\documentclass[aps,prl,twocolumn,superscriptaddress,longbibliography]{revtex4-1}
\usepackage{graphicx}
\usepackage{amssymb,amsmath}
\usepackage{bm}
\usepackage{dcolumn}
\usepackage{subfigure}
\usepackage{float}
\usepackage[OT1]{fontenc} 
\usepackage{url}
\usepackage{slashed}
\usepackage{color}
\usepackage{verbatim}
\usepackage{times}
\usepackage{graphicx}
\usepackage[colorlinks,linkcolor=blue,anchorcolor=blue,citecolor=blue]{hyperref}
\usepackage{ dsfont }

\begin{document}
	
\title{ Randomness-enhanced expressivity of quantum neural networks}

\author{Yadong Wu}
\affiliation{Department of Physics, Fudan University, Shanghai, 200438, China}
\affiliation{State Key Laboratory of Surface Physics, Key Laboratory of Micro and Nano Photonic Structures (MOE), Institute for Nanoelectronic Devices and Quantum Computing, Fudan University, Shanghai 200438, China}
\affiliation{Shanghai Qi Zhi Institute, AI Tower, Xuhui District, Shanghai 200232, China}

\author{Juan Yao}
\affiliation{Shenzhen Institute for Quantum Science and Engineering, Southern University of Science and Technology, Shenzhen 518055, Guangdong, China}
\affiliation{International Quantum Academy, Shenzhen 518048, Guangdong, China}
\affiliation{Guangdong Provincial Key Laboratory of Quantum Science and Engineering, Southern University of Science and Technology, Shenzhen 518055, Guangdong, China}

\author{Pengfei Zhang}
\email{pengfeizhang.physics@gmail.com}
\affiliation{Department of Physics, Fudan University, Shanghai, 200438, China}
\affiliation{Shanghai Qi Zhi Institute, AI Tower, Xuhui District, Shanghai 200232, China}

\author{Xiaopeng Li}
\email{xiaopeng$_$li@fudan.edu.cn}
\affiliation{Department of Physics, Fudan University, Shanghai, 200438, China}
\affiliation{State Key Laboratory of Surface Physics, Key Laboratory of Micro and Nano Photonic Structures (MOE), Institute for Nanoelectronic Devices and Quantum Computing, Fudan University, Shanghai 200438, China}
\affiliation{Shanghai Qi Zhi Institute, AI Tower, Xuhui District, Shanghai 200232, China}
\affiliation{ Shanghai Artificial Intelligence Laboratory, Shanghai 200232, China} 
\affiliation{Shanghai Research Center for Quantum Sciences, Shanghai 201315, China}
	
\begin{abstract}
As a hybrid of artificial intelligence and quantum computing, quantum neural networks (QNNs) have gained significant attention as a promising application on near-term, noisy intermediate-scale quantum (NISQ) devices. Conventional QNNs are described by parametrized quantum circuits, which perform unitary operations and measurements on quantum states. In this work, we propose a novel approach to enhance the expressivity of QNNs by incorporating randomness into quantum circuits. Specifically, we introduce a random layer, which contains single-qubit gates sampled from an trainable ensemble pooling. The prediction of QNN is then represented by an ensemble average over a classical function of measurement outcomes. We prove that our approach can accurately approximate arbitrary target operators using Uhlmann's theorem for majorization, which enables observable learning. Our proposal is demonstrated with extensive numerical experiments, including observable learning, R\'enyi entropy measurement, and image recognition. We find the expressivity of QNNs is enhanced by introducing randomness for multiple learning tasks, which could have broad application in quantum machine learning.
\end{abstract}
	
\maketitle

\emph{ \color{blue}Introduction.--} 
In recent years, significant breakthroughs have been made in the field of artificial intelligence. Among various machine learning algorithms, neural networks have played a vital role, thanks to their universal expressivity for deep architectures. As a quantum generalization of neural networks, quantum neural networks (QNNs) have been proposed based on parameterized quantum circuits. QNNs use quantum states instead of classical numbers as inputs\cite{PhysRevA.101.032308,PhysRevA.98.062324,farhi2018classification,Benedetti_2019}. However, the evolution of the input quantum states is constrained to be unitary, which limits the expressivity of QNNs. For physical observables, which are linear functions of the input quantum states or density matrices, QNNs can achieve high accuracy only if the target operator shares the same eigenvalues with the measurement operator. For a general situation, it requires introducing auxiliary qubits, as proposed in \cite{PhysRevResearch.3.L032049}. To expresse non-linear functions of the input density matrices, such as purities, traditional approaches introduce multiple replicas, which is unfavorable on near-term, noisy intermediate-scale quantum (NISQ) devices with a limited number of logical qubits. Previous studies have also reported moderate accuracy for more general machine learning tasks, including image recognization \cite{PhysRevResearch.3.L032057,PRXQuantum.2.040321,Caro:2022vx,Abbas:2021wp,PhysRevA.103.032430}.

    \begin{figure}[t]
        \centering
        \includegraphics[width=0.9\linewidth]{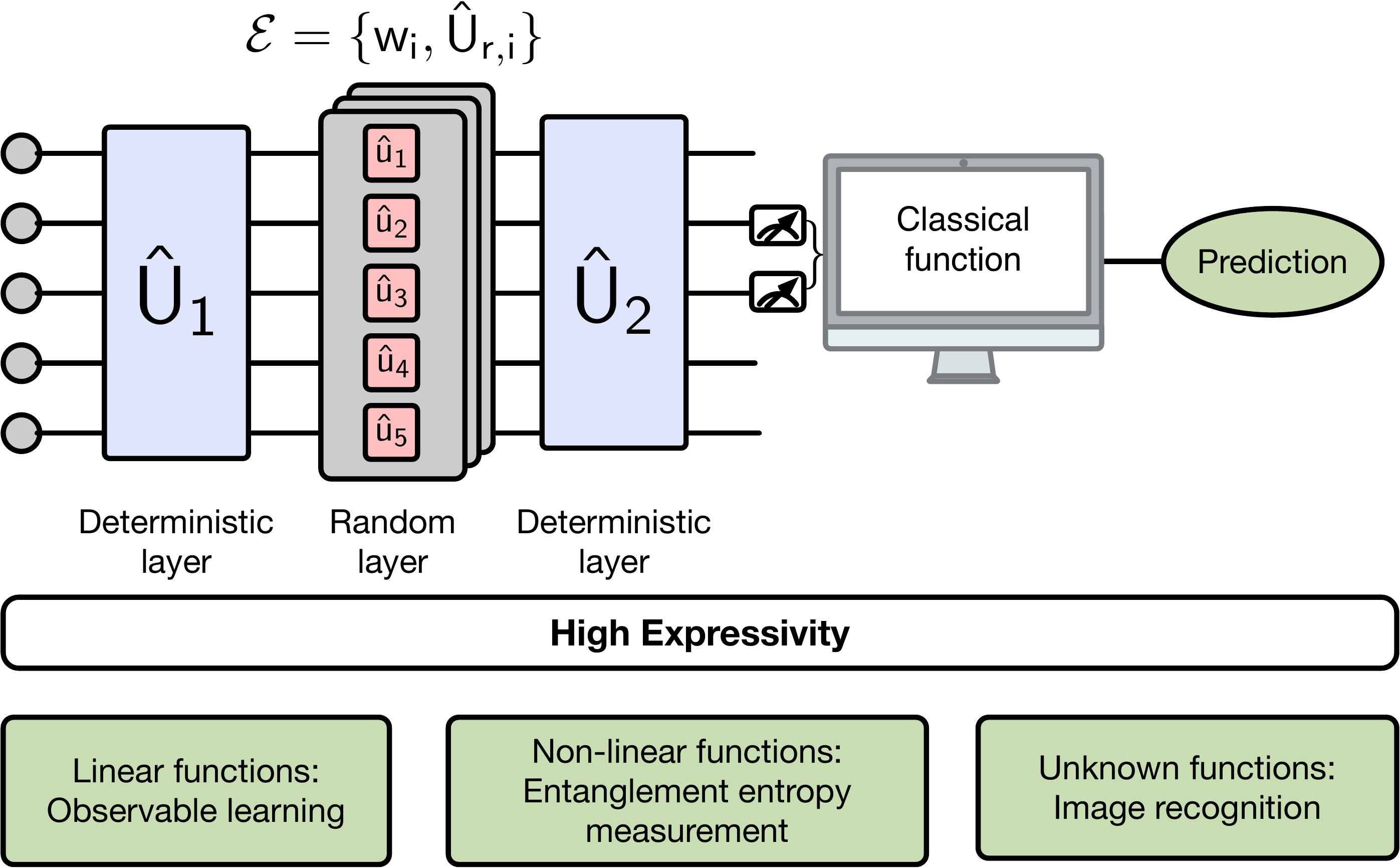}
        \caption{ \textit{An illustration is provided for the proposed architecture of randomized quantum neural networks.} In this example, the circuit contains two deterministic layers $\hat{U}_{1(2)}$ and one random layer $\hat{U}_r$ in between, with the final measurement performed on two qubits. As demonstrated in this work, this architecture shows randomness-enhanced expressivity for a variety of general learning tasks. }
        \label{fig:schemticas}
    \end{figure}

In this work, we propose a universal scheme to overcome the expressivity obstacle without the need for additional replicas. Our main inspiration comes from the recent development of the randomized measurement toolbox for quantum simulators \cite{Qi:2019rpi,2017arXiv171101053A,2017arXiv171101053A,PhysRevLett.122.120505,PhysRevLett.120.050406,PhysRevLett.108.110503,PhysRevA.99.052323,Knips:2020uu,Huang:2020wo,Elben:2023tg,PhysRevResearch.4.013054,PRXQuantum.2.030348,2021arXiv210505992A,2021arXiv211002965L,PhysRevLett.127.110504,2022arXiv220713723W,doi:10.1126/science.abk3333,2022arXiv220203272B,2023PhRvA.107d2403K,2023ScPP...14...94S,PhysRevLett.130.230403,2022arXiv220912924B,2022arXiv221109835A,Akhtar:2022nme,PhysRevResearch.5.023027,koh2022classical,koh2022classical}. In all of these protocols, a measurement is performed after a random unitary gate, and the desired property is predicted through a classical computer after collecting sufficient measurement outcomes. In particular, the random measurement has been experimentally realized in \cite{Vovrosh:2021tc,PhysRevResearch.3.043122,Noel:2022wr,koh2022experimental,doi:10.1126/science.aau4963,PRXQuantum.2.010307}. These developments unveil that randomness plays a central role in extracting information from complex quantum systems efficiently. From a machine learning perspective, this implies that introducing random unitaries can enhance the expressivity of QNNs. This naturally leads to the concept of randomized quantum neural networks, where we collect measurement outcomes from an ensemble of parametrized quantum circuits to make final predictions. Analogous to the different types of layers in classical neural networks, randomized QNNs consist of deterministic layers and random layers. In deterministic layers, the quantum gates contain parameterized quantum gates as in traditional QNNs, while in random layers, they are sampled from trainable ensembles of single-qubit gates. This is illustrated in FIG. \ref{fig:schemticas}. We demonstrate the high expressivity of the proposed architecture using several different tasks, including both linear and nonlinear functions of the input density matrix. Our results pave the way towards realizing the universal expressivity ability for QNNs.
\vspace{5pt}

\emph{ \color{blue}Architecture.--} 
We begin with a detailed description of randomized QNNs. To be concrete, we focus on the architecture illustrated in FIG. \ref{fig:schemticas} for $N_{\text{sys}}=5$ qubits, which comprises two deterministic layers, namely $\hat{U}_1$ and $\hat{U}_2$, with a random single qubit gate layer $\hat{U}_r$ in between. 

Each deterministic layer $\hat{U}_{l_d}$ ($l_d=1,2$) contains a number of units $\hat{V}_{l_d}^{l}({\bm \theta}_{l_d}^l)$ ($l\in \{1,2,...,L_{l_d}\}$) and each deterministic layer is constructed as 
\begin{equation}
\hat{U}_{l_d}=\hat{V}_{l_d}^{L_{l_d}}(\bm{\theta}_{l_d}^{L_{l_d}})...\hat{V}_i^{2}(\bm{\theta}_{l_d}^2)\hat{V}_{l_d}^{1}(\bm{\theta}_{l_d}^1),
\end{equation}
where $\{\bm{\theta}_{l_d}^l\}$ are the parameters of the deterministic layers. In general, the arrangement of two-qubit gates in each deterministic layer allows for a large degree of freedom. In this work, we focus on the standard brick wall architecture with spatial locality. Each unit $\hat{V}_{l_d}^l$ contains $N_{\rm sys}-1$ two qubit gates and each two qubit gate is a SU(4) matrix which can be parameterized as $\exp(\sum_jc_j\hat{g}_j)$. Here $\hat{g}_j$ is the generator of SU(4) group and $\{\bm{\theta}_{l_d}^l\}$ denotes parameters $\{\mathbf{c}\}$ of all two qubit gates \cite{supp}. Nonetheless, alternative choices for each deterministic layer have the potential to enhance the expressivity of QNNs for a fixed number of gates \cite{PhysRevResearch.3.L032057}.

For the sake of experimental convenience, the random layer $\hat{U}_r$ comprises a tensor product of single-qubit gates, denoted as $\hat{u}_1\otimes \hat{u}_2... \otimes \hat{u}_{N_{\text{sys}}}$. These gates are sampled from an ensemble
\begin{equation}
\mathcal{E}=\{(w_i,\hat{U}_{r,i}=\hat{u}_1^i(\bm{\alpha}_{i}^{1})\otimes \hat{u}_2^i(\bm{\alpha}_{i}^{2})... \otimes \hat{u}_{N_{\text{sys}}}^i(\bm{\alpha}_{i}^{{N_{\text{sys}}}}))\},
\end{equation} 
where $i=1,2,...,N_r$ labels different elements and $w_i$ is the corresponding weight with $\sum_i w_i=1$. Each single qubit gate is parametrized by generators of SU(2) with 3 dimensional real vector $\bm{\alpha}_{i}^{q}$ ($q\in \{1,2,...,{N_{\text{sys}}}\}$). Both $\{w_i\}$ and $\{\bm{\alpha}_{i}^q\}$ are trainable parameters. It is also straightforward to introduce multiple random layers into the full architecture of QNNs. Importantly, it is worth noting the differences between our definition and typical random measurement protocols. Firstly, our random layer can be added at any point in the quantum circuit, not necessarily before the final measurement. Secondly, our definition of $\mathcal{E}$ allows for non-trivial correlations between single-qubit gates on different sites, which is typically absent in random measurement protocols. Both features are necessary for achieving a high expressivity in QNNs.

We consider a dataset $\{(|\psi_m\rangle, \mathcal{T}_m)\}$, in which $m\in\{1,2,...,N_D\}$ labels different data and $\mathcal{T}_m$ is the target information for the corresponding state $|\psi_m\rangle$. For each unitary $\hat{U}_{r,i}$ in the ensemble $\mathcal{E}$, we perform projective measurements in the computational basis for $k\sim O(1)$ qubits. The small number of measured qubits would avoid the barren plateaus, which can be caused by global measurements \cite{cerezo2021cost}.  In FIG. \ref{fig:schemticas}, we set $k=2$, and the measurement yields the probability distribution given by:
\begin{equation}
p^{ss'}_{i,m}=\langle \psi_m| \hat{U}_1^\dagger\hat{U}_{r,i}^\dagger\hat{U}_2^\dagger(\hat{P}_s^2\otimes\hat{P}_{s'}^3)\hat{U}_2\hat{U}_{r,i}\hat{U}_1|\psi_m\rangle,
\end{equation}
where the projection operator $\hat{P_s^q}=\frac{1+s\hat{\sigma}_z^q}{2}$ for $s=\pm 1$. Due to the constraint $\sum_{ss'} p^{ss'}_{i,m}=1$, there are only $3$ non-trivial components of $p^{ss'}_{i,m}$, denoted by the vector $\mathbf{p}_{i,m}$. We then use a classical computer to apply a general function $f_{\bm{\beta}}(.)$, parametrized by $\bm{\beta}$, to the probability distribution $p^{ss'}_{i,m}$, which yields a single outcome denoted by $\mathcal{P}_{i,m}=f_{\bm{\beta}}(\mathbf{p}_{i,m})$. The classical function can be described by elementary functions in the simplest setting, but is more generally described by classical neural networks. We further average the outcome over the ensemble $\mathcal{E}$ to obtain the final prediction for the input state $|\psi_m\rangle$ as:
\begin{equation}
\mathcal{P}_m=\sum_{i=1}^{N_r} w_i \mathcal{P}_{i,m}=\sum_{i=1}^{N_r} w_i f_{\bm{\beta}}(\mathbf{p}_{i,m}).
\end{equation}
We use the mean square error (MSE) as the loss function $\mathcal{L}=\frac{1}{N_D}\sum_{m}(\mathcal{P}_m-\mathcal{T}_m)^2$ with a data size of $N_D$ during the training process. We apply the gradient descent algorithm to optimize the parameters $\{\bm{\theta}_{l_d}^l, w_i, \bm{\alpha}^q_i, \bm{\beta}\}$ to minimize the loss function $\mathcal{L}$, and set the numerical criteria as $\mathcal{L}<10^{-5}$ to characterize the accurate prediction. Our method to compute gradients of parameters is explained in the Supplementary Material \cite{supp}. In the following sections, we focus on demonstrating high expressivity for randomized QNNs. Our examples range from simple physical tasks including observable learning and R\'enyi entropy measurement, to standard machine learning tasks such as image recognization. 

\vspace{5pt}
\emph{ \color{blue}Observable learning.--} 
To show the high expressivity of randomized QNNs, let us consider a simple scenario where the target, $\mathcal{T}_m$, is an expectation of a physical observable $\hat{O}$ with $\mathcal{T}_m=\langle \psi_m|\hat{O}|\psi_m\rangle$. For simplicity, focusing on single-qubit measurement with $k=1$, we first investigate whether the randomized QNNs as proposed as in FIG. \ref{fig:schemticas} can approximate the target function $\mathcal{T}_m$ as accurate as possible for sufficiently deep circuit structures with sufficiently large $N_r$. As physical observables are linear in density matrices, a linear function $f_{\bm{\beta}}(x)=\beta_0+\beta_1 x$ will be applied to the measurement result. Explicitly, we introduce $\hat{U}_{\text{tot},i}^{}$ for a random realization $i$ of the quantum circuit. As an example, we have $\hat{U}_{\text{tot},i}^{}=\hat{U}_2\hat{U}_{r,i}\hat{U}_1$. An accurate prediction of the target function requires that 
\begin{equation}\label{eqn:obs_requirement}
\sum_{i=1}^{N_r} w_i~\hat{U}_{\text{tot},i}^\dagger (\beta_0\hat{\sigma}_0^1+\beta_1\hat{\sigma}_z^1) \hat{U}_{\text{tot},i}^{}=\hat{O},
\end{equation}
where $\hat{\sigma}_0$ is the identity operator and Pauli matrix $\hat{\sigma}_z$ is the single-qubit's measurement operator.
 \begin{figure}[t]
        \centering
        \includegraphics[width=0.85\linewidth]{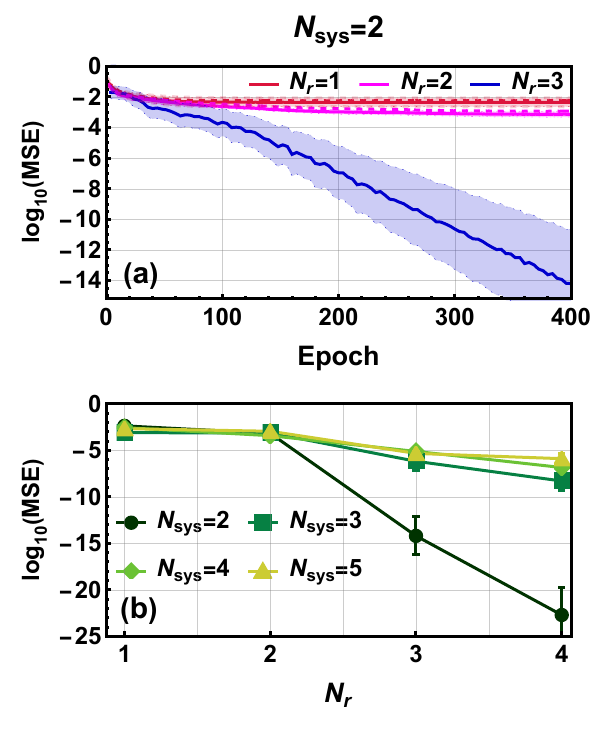}
        \caption{
        { \textit{Predicting observables using QNN with a random layer.}} 
         (a) The logarithmic training mean square error is shown as a function of the training epoch for observable learning with $N_{\text{sys}}=2$. The solid lines represent the averaged over the training process for 10 different random target operators with independent runs, while the shaded region represents the standard deviation. The dashed lines are the validation loss with the dataset containing 200 samples. 
          (b) The logarithmic mean square error of training dataset for $N_{\rm sys}\in\{2,3,4,5\}$ and $N_r\in\{1,2,3,4\}$. The markers represents the average over 10 different random target operators with random initializations, and error bars are the standard deviation.  }
        \label{fig:obs}
    \end{figure}

For the case of $N_r=1$ and $w_1=1$, our setup reduces to the traditional QNN without randomness. In this scenario, Eq.\eqref{eqn:obs_requirement} requires that $(\beta_0\hat{\sigma}_0^1+\beta_1\hat{\sigma}_z^1)$ and $\hat{O}$ be related by a unitary transformation. Since the unitary transformation preserves the eigenvalues of the operator, the requirement cannot be satisfied for a general operator $\hat{O}$. When $N_r>1$, Eq.\eqref{eqn:obs_requirement} can be expressed as $\Phi(\hat{\Sigma})=\hat{O}$, where $\hat{\Sigma}\equiv \beta_1\hat{\sigma}_z^1+\beta_0\hat{\sigma}_0^1$ and $\Phi(\hat{X})$ is a mixed-unitary channel \cite{watrous_2018}. For sufficiently complex circuit structures, we expect $\Phi$ to be generic. In comparison to the $N_r=1$ case, there is no constraint from unitarity. However, we still need to ask whether Eq.\eqref{eqn:obs_requirement} can be satisfied for an arbitrary operator $\hat{O}$. In the following, we prove that the answer to this question is affirmative:

\textbf{Step 1.} Mathematically, if there exists a mixed-unitary channel $\Phi$ such that $Y=\Phi(X)$, we say that $X$ majorizes $Y$, denoted by $Y\prec X$ \cite{nielsen2002introduction}. Thus, for a randomized QNNs which can accurately predict any observable $\hat{O}$, we need to find values of $\beta_0$ and $\beta_1$ such that $ \hat{O} \prec \hat{\Sigma}$ for any $\hat{O}$.

\textbf{Step 2.} According to Uhlmann's theorem for majorization \cite{Uhlmann,nielsen2002introduction}, $ \hat{O} \prec \hat{\Sigma}$ if and only if $ \bm{\lambda}_{\hat{O}} \prec \bm{\lambda}_{\hat{\Sigma}}$, where $\bm{\lambda}_{\hat{X}}$ is the list of eigenvalues for the operator $\hat{X}$ in descending order. Here the majorization between two real vectors $\bm{y} \prec \bm{x}$ is defined as  (i) $\sum_{j=1}^qx_j \geq \sum_{j=1}^qy_j$ for arbitrary $1\leq q <\mathcal{D}$  and (ii) $\sum_{j=1}^\mathcal{D}x_j = \sum_{j=1}^\mathcal{D}y_j$. Here $\mathcal{D}$ is the dimension of the vectors. Noting that condition (ii) takes into account the trace-preserving property of mixed-unitary channels.

\textbf{Step 3.} We can always find $\beta_0$ and $\beta_1$ such that $\bm{\lambda}_{\hat{O}} \prec \bm{\lambda}_{\hat{\Sigma}}$. Assuming $\beta_1>0$, the first or last $\mathcal{D}/2$ components of $\bm{\lambda}_{\hat{\Sigma}}$ correspond to the values $\beta_0+\beta_1$ or $\beta_0-\beta_1$, respectively. The constant term $\beta_0$ can then be determined using condition (ii), which gives $\beta_0=\mathcal{D}^{-1}\sum_{j=1}^\mathcal{D} \lambda_{\hat{O},j}$. Moreover, condition (i) can always be satisfied for sufficiently large $\beta_1$. This proves the existence of $\beta_0$ and $\beta_1$ such that $ \hat{O} \prec \hat{\Sigma}$.
\vspace{3pt}

Although randomized QNNs have the potential to express arbitrary operators, it is difficult to  determine an upper bound  or a required value for $N_r$ in practical learning tasks. It is unfavorable to have large $N_r$ or a large number of random layers, especially in NISQ devices. Therefore, we turn to numerical simulations of the randomized QNNs, and investigate practical requirements on $N_r$. Since the basis change can be efficiently captured by the deterministic layer $\hat{U}_1$, we focus on observables $\hat{O}$ that are diagonal in the computational basis. For simplicity, we further set $\hat{U}_1=\hat{I}$ and $\hat{U}_2$ composed by $L_2$ units of a brick wall structure \cite{supp}. For each system size ${N_{\text{sys}}}$, we test whether a random diagonal operator $\hat{O}$ can be predicted accurately for different values of $N_r$ by monitoring the training loss for a sufficiently large dataset. As an example, we plot the logarithmic training mean square error $\rm \log_{10}(MSE)$ as a function of the training epoch for ${N_{\text{sys}}}=2$ in FIG. \ref{fig:obs} (a). The curves are averaged over 10 operators with random eigenvalues from the uniform distribution $[-2.5,2.5]$. When we increase $N_r$ from 1 to 3, there is a rapid decrease in the training loss for large training epochs. The result shows that $N_r=3$ is sufficient for learning general operators for ${N_{\text{sys}}}=2$ where the loss $\mathcal{L}$ can be decreased to $10^{-14}$. We further extend the system size ${N_{\text{sys}}}$ to study how it affects the number of required random gates. The results are shown in FIG. \ref{fig:obs} (b). Although we are limited to small system sizes ${N_{\text{sys}}} \in \{2,3,4,5\}$, the results clearly show weak dependence of $N_r$ on $N_{\text{sys}}$. The training results show that $N_r=3$ already gives highly accurate predictions for ${N_{\text{sys}}}=5$. 

\begin{figure}[t]
        \centering
        \includegraphics[width=0.98\linewidth]{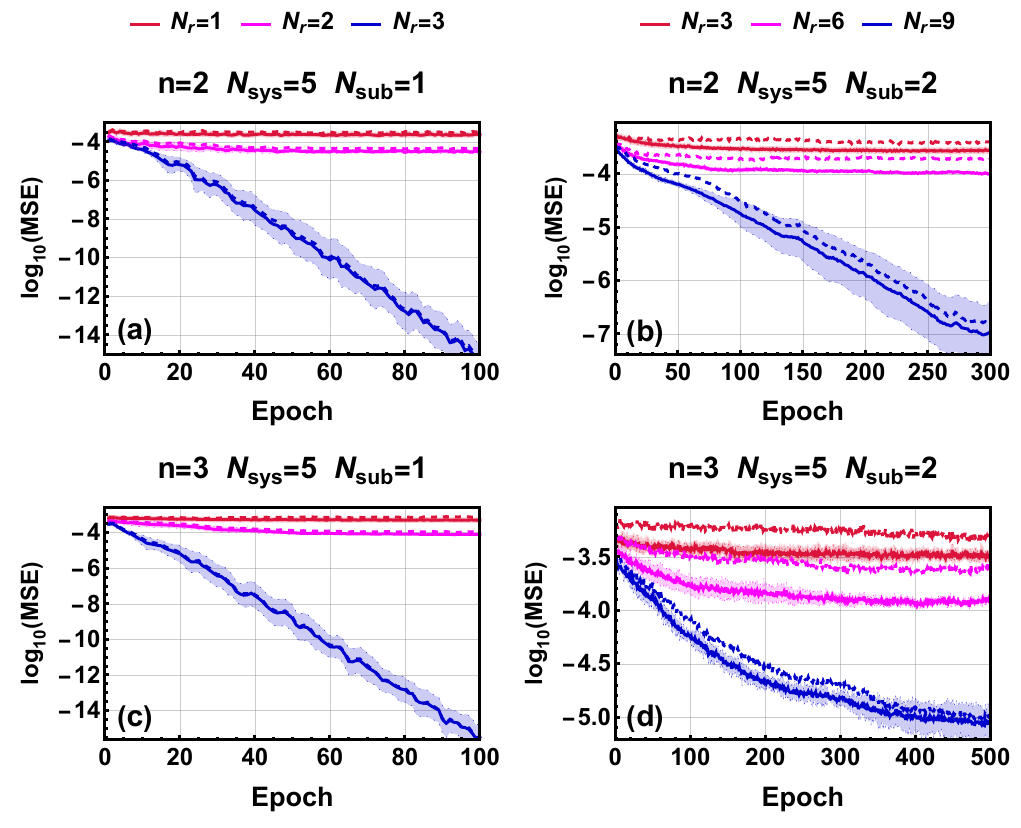}
        \caption{
        \textit{Predicting R\'enyi entropies using QNN with a random layer. } The logarithmic training mean square error is shown as a function of the training epoch for purity with ${N_{\text{sys}}}=5$ and (a) $n=2, {N_{\text{sub}}}=1$,  (b) $n=2, {N_{\text{sub}}}=2$, (c) $n=3, {N_{\text{sub}}}=1$ or (d) $n=3, {N_{\text{sub}}}=2$. The results are averaged over the training process for 10 different random initializations, and the shaded region represents the standard deviation. The dashed lines are the validation loss with the dataset containing 200 samples.  }
        \label{fig:Renyi}
    \end{figure}

\vspace{5pt}
\emph{ \color{blue}R\'enyi entropy measurement.--} 
We now consider targets that are non-linear functions of density matrices. One example is the R\'enyi entropy, which is also of experimental interest. To compute the R\'enyi entropy for a subsystem $A$ consisting of the central $N_{\text{sub}}$ qubits, we first calculate the reduced density matrix $\hat{\rho}_A$ of an input state $|\psi_m\rangle$ by tracing out the degrees of freedom of the complementary subsystem $\bar{A}$. We then select the target as
\begin{equation}
\mathcal{T}_m=\text{Tr}_A[\hat{\rho}_A^n],
\end{equation}
which is related to the n-th R\'enyi entropy through $S_{A}^{(n)}=-\frac{1}{n-1} \ln(\mathcal{T}_m)$. Since we are directly measuring a local property of the input wavefunction, it is reasonable to fix $\hat{U}_1$ and $\hat{U}_2$ to the identity matrix $\hat{I}$ and focus on the random layer $\hat{U}_r$ with $k=N_{\text{sub}}$. This approach provides a minimum guaranteed expressivity of randomized QNNs. Because the target $\mathcal{T}_m$ is proportional to $\rho^n$, we choose the function $f_{\bm{\beta}}(\mathbf{x})$ to be a polynomial up to the $n$-th order. However, it is worth noting that lower order polynomials may also work in certain cases \cite{Renyi3Poly}. We prepare a dataset with random states $|\psi_m\rangle$, the detailed description of which is provided in the Supplementary Material \cite{supp}. The numerical results for $n=2,3$, $N_\text{sys}=5$ and ${N_{\text{sub}}}=1,2$ are shown in FIG. \ref{fig:Renyi}. To achieve accurate predictions, we need $N_r=3$ for ${N_{\text{sub}}}=1$ and $N_r=9$ for ${N_{\text{sub}}}=2$. The blue lines in FIG. \ref{fig:Renyi} demonstrate that the loss $\mathcal{L}$ is able to reach a value of $10^{-5}$ and still keep decreasing, indicating the ability to make accurate predictions. We have also discussed the saturation of $N_r$ for $n=2, N_{\rm sub}=2$ and the required number of $N_r$ if we instead consider $n=3$. The results are shown in the Supplementary Material \cite{supp}. 

It is interesting to compare our results to the proposed random measurement protocol for R\'enyi entropies. Our results indicate that $N_r$ scales as $3^{N_{\text{sub}}}$ when measuring R\'enyi entropies. In comparison, the previous protocol required each single-qubit gate $\hat{u}_q^i$ to be sampled from the circular unitary ensemble \cite{doi:10.1126/science.aau4963,Huang:2020wo}. For $n=2$, the circular unitary ensemble can be replaced by unitary 2-designs, which are known to be the Clifford group. Since the single-qubit Clifford group contains 24 elements, the total number of unitary matrices $\hat{U}_{r,i}$ would naively scale as $24^{N{\text{sub}}}$. However, in practice, this can be significantly reduced because randomized measurement protocols only require $N_\text{s}$ snapshots sampled from the full ensemble. The theoretical bound of $N_\text{s}$ for measuring general linear observables in a subsystem with $N_{\text{sub}}$ qubits using random Pauli measurements up to an error $\epsilon$ is given by $N_\text{s} \gtrsim 3^{N_{\text{sub}}}/\epsilon^2$ \cite{Elben:2023tg,Huang:2020wo}. Consequently, in this quantum neural network structure, the number of unitary matrices $\hat{U}_{r,i}$ that contribute is at most $3^{N_{\text{sub}}}$, as in our randomized QNNs.

\begin{figure}[t]
        \centering
        \includegraphics[width=0.75\linewidth]{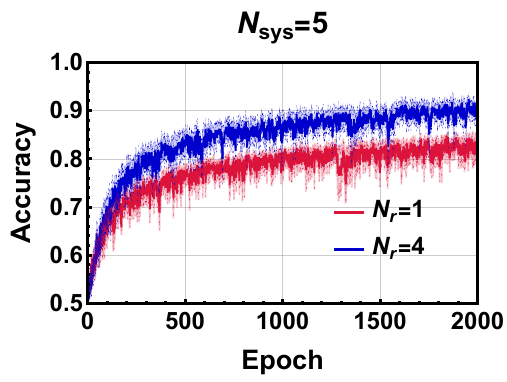}
        \caption{  \textit{Image recognition  by QNN with a random layer. }The accuracy is shown as a function of the training epoch for the image recognition task. The results are averaged over the training process for 10 different random initializations, and the shaded region represents the standard deviation.   }
        \label{fig:image}
    \end{figure}

\vspace{5pt}
\emph{ \color{blue}Image recognition.--} 
Finally, we turn our attention to image recognition, a more practical machine learning task, in order to demonstrate the enhanced expressivity of randomized QNNs. In this case, we use Google's 'Street View of House Number (SVHN)' dataset as an example \cite{netzer2011reading}. Each image in the dataset corresponds to an integer number. For demonstration purposes, we select two categories of images containing the numbers '1' and '4'. Initially, we compress each image into an $8\times 8$ pixel format, resulting in a $64$-dimensional real vector, which can be equivalently represented as a $32$-dimensional complex vector. Subsequently, we encode the image into the input wave function using $N_{\text{sys}}=5$ qubits \cite{supp}. Unlike previous tasks, the mapping between the input and the output is highly complex and non-local, lacking a simple understanding. Consequently, we allow both $\hat{U}_1$ and $\hat{U}_2$ to be trainable. After measuring a single qubit, we choose a 5$^{\rm th}$-order polynomial for the function $f_{\bm{\beta}}(\bm{x})$. Since the image recognition is a two category classification task, after obtaining the final ensemble average prediction $\mathcal{P}_m$, we apply a logistic-sigmoid function to restrict the prediction in the region (0,1) with $\mathcal{G}_m=1/(1+\exp(-\mathcal{P}_m))$. And we use the cross-entropy as the loss function $\mathcal{L}=\frac{1}{N_D}\sum_m-\mathcal{T}_m\log(\mathcal{G}_m)-(1-\mathcal{T}_m)\log(1-\mathcal{G}_m)$ to optimize the parameters in the randomized QNN. The accuracy $F=\frac{1}{N_D}\sum_m|[{\rm sign}(\mathcal{G}_m-0.5)+1]/2-\mathcal{T}_m|$ for $N_r=1$ and $N_r=4$ are shown in FIG. \ref{fig:image}. For $N_r=1$, the accuracy saturates at approximately 0.8 after a large number of epochs, while the averaged accuracy for the test dataset reaches $69.8\%$. The introduction of a single random layer with $N_r=4$ significantly enhances the accuracy of the predictions. In this case, the training dataset achieves an accuracy higher than $90\%$, and the average accuracy for the test dataset is $82.29\%$. The utilization of a non-trivial random layer with $N_r=4$ demonstrates a significant improvement in the prediction capabilities of QNNs, indicating the enhanced expressivity of our randomized QNN architecture.

\vspace{5pt}
\emph{ \color{blue}Outlook.--} 
This work introduces the concept of randomized quantum neural networks, which include random layers where quantum gates are selected from an ensemble of unitary matrices. It is proven that these random layers provide universal expressivity for general physical observables using Uhlmann's theorem for majorization. Numerical simulations further show that this architecture achieves high expressivity for non-linear functions of the density matrix, such as R\'enyi entropies and image recognition, with small ensemble sizes $N_r$. These results indicate that the proposed method has potential for broad applications in NISQ devices. We remark that adding a random layer to the QNNs causes an extra computational cost proportional to $N_r$. Nonetheless, introducing randomness into QNNs while maintaining the same computational cost still improves the learning performance significantly~\cite{supp}. We further highlight the differences between our architecture and the proposal presented in a very recent paper \cite{Xiaoting}. Their work also incorporates a series of parameterized quantum circuits, where the circuit consists of multiple parametrized (controlled-) rotations that share the same parameter. In contrast, our architecture features only a few random layers described by a tensor product of single-qubit gates, making its training process more efficient. 

While the focus of this work is on parameterized quantum circuits with brick wall structures, it is straightforward to combine this novel architecture with other proposals to further improve expressivity or learning efficiency. For instance, it is possible to add ancilla qubits and explore more sophisticated architectures for the deterministic layers. Additionally, it would be interesting to investigate the impact of random layers on other quantum machine learning algorithms beyond traditional quantum neural networks \cite{romero2017quantum,bondarenko2020quantum,oh2020tutorial,hur2022quantum,cong2019quantum,Caro:2022aa}, such as quantu autoencoders \cite{romero2017quantum,bondarenko2020quantum}.

\vspace{5pt}
 \emph{\color{blue}Acknowledgement.--} 
 We thank Yingfei Gu, Ning Sun, Ce Wang, Hai Wang, and Yi-Zhuang You for helpful discussions. 
YW and XL are supported by National Program on Key Basic Research Project of China (Grant No. 2021YFA1400900), National Natural Science Foundation of China (Grants No. 11934002), Shanghai Municipal Science and Technology Major Project (Grant No. 2019SHZDZX01). 
YW is supported by the National Natural Science Foundation of China (Grant No. 12174236).
JY is supported by the National Natural Science Foundation of China (Grant No. 11904190).

\bibliographystyle{unsrt}  
\bibliography{BibRandom}

\begin{thebibliography}{10}

\bibitem{PhysRevA.101.032308}
Maria Schuld, Alex Bocharov, Krysta~M. Svore, and Nathan Wiebe.
\newblock Circuit-centric quantum classifiers.
\newblock {\em Phys. Rev. A}, 101:032308, Mar 2020.

\bibitem{PhysRevA.98.062324}
Jin-Guo Liu and Lei Wang.
\newblock Differentiable learning of quantum circuit born machines.
\newblock {\em Phys. Rev. A}, 98:062324, Dec 2018.

\bibitem{farhi2018classification}
Edward Farhi and Hartmut Neven.
\newblock Classification with quantum neural networks on near term processors,
  2018.

\bibitem{Benedetti_2019}
Marcello Benedetti, Erika Lloyd, Stefan Sack, and Mattia Fiorentini.
\newblock Parameterized quantum circuits as machine learning models.
\newblock {\em Quantum Science and Technology}, 4(4):043001, nov 2019.

\bibitem{PhysRevResearch.3.L032049}
Yadong Wu, Juan Yao, Pengfei Zhang, and Hui Zhai.
\newblock Expressivity of quantum neural networks.
\newblock {\em Phys. Rev. Res.}, 3:L032049, Aug 2021.

\bibitem{PhysRevResearch.3.L032057}
Yadong Wu, Pengfei Zhang, and Hui Zhai.
\newblock Scrambling ability of quantum neural network architectures.
\newblock {\em Phys. Rev. Res.}, 3:L032057, Sep 2021.

\bibitem{PRXQuantum.2.040321}
Leonardo Banchi, Jason Pereira, and Stefano Pirandola.
\newblock Generalization in quantum machine learning: A quantum information
  standpoint.
\newblock {\em PRX Quantum}, 2:040321, Nov 2021.

\bibitem{Caro:2022vx}
Matthias~C. Caro, Hsin-Yuan Huang, M.~Cerezo, Kunal Sharma, Andrew Sornborger,
  Lukasz Cincio, and Patrick~J. Coles.
\newblock Generalization in quantum machine learning from few training data.
\newblock {\em Nature Communications}, 13(1):4919, 2022.

\bibitem{Abbas:2021wp}
Amira Abbas, David Sutter, Christa Zoufal, Aurelien Lucchi, Alessio Figalli,
  and Stefan Woerner.
\newblock The power of quantum neural networks.
\newblock {\em Nature Computational Science}, 1(6):403--409, 2021.

\bibitem{PhysRevA.103.032430}
Maria Schuld, Ryan Sweke, and Johannes~Jakob Meyer.
\newblock Effect of data encoding on the expressive power of variational
  quantum-machine-learning models.
\newblock {\em Phys. Rev. A}, 103:032430, Mar 2021.

\bibitem{Qi:2019rpi}
Xiao-Liang Qi, Emily~J. Davis, Avikar Periwal, and Monika Schleier-Smith.
\newblock {Measuring operator size growth in quantum quench experiments}.
\newblock 6 2019.

\bibitem{2017arXiv171101053A}
Scott {Aaronson}.
\newblock {Shadow Tomography of Quantum States}.
\newblock {\em arXiv e-prints}, page arXiv:1711.01053, November 2017.

\bibitem{PhysRevLett.122.120505}
Andreas Ketterer, Nikolai Wyderka, and Otfried G\"uhne.
\newblock Characterizing multipartite entanglement with moments of random
  correlations.
\newblock {\em Phys. Rev. Lett.}, 122:120505, Mar 2019.

\bibitem{PhysRevLett.120.050406}
A.~Elben, B.~Vermersch, M.~Dalmonte, J.~I. Cirac, and P.~Zoller.
\newblock R\'enyi entropies from random quenches in atomic hubbard and spin
  models.
\newblock {\em Phys. Rev. Lett.}, 120:050406, Feb 2018.

\bibitem{PhysRevLett.108.110503}
S.~J. van Enk and C.~W.~J. Beenakker.
\newblock Measuring $\mathrm{Tr}{\ensuremath{\rho}}^{n}$ on single copies of
  $\ensuremath{\rho}$ using random measurements.
\newblock {\em Phys. Rev. Lett.}, 108:110503, Mar 2012.

\bibitem{PhysRevA.99.052323}
A.~Elben, B.~Vermersch, C.~F. Roos, and P.~Zoller.
\newblock Statistical correlations between locally randomized measurements: A
  toolbox for probing entanglement in many-body quantum states.
\newblock {\em Phys. Rev. A}, 99:052323, May 2019.

\bibitem{Knips:2020uu}
Lukas Knips, Jan Dziewior, Waldemar K{\l}obus, Wies{\l}aw Laskowski, Tomasz
  Paterek, Peter~J. Shadbolt, Harald Weinfurter, and Jasmin D.~A. Meinecke.
\newblock Multipartite entanglement analysis from random correlations.
\newblock {\em npj Quantum Information}, 6(1):51, 2020.

\bibitem{Huang:2020wo}
Hsin-Yuan Huang, Richard Kueng, and John Preskill.
\newblock Predicting many properties of a quantum system from very few
  measurements.
\newblock {\em Nature Physics}, 16(10):1050--1057, 2020.

\bibitem{Elben:2023tg}
Andreas Elben, Steven~T. Flammia, Hsin-Yuan Huang, Richard Kueng, John
  Preskill, Beno{\^\i}t Vermersch, and Peter Zoller.
\newblock The randomized measurement toolbox.
\newblock {\em Nature Reviews Physics}, 5(1):9--24, 2023.

\bibitem{PhysRevResearch.4.013054}
Hong-Ye Hu and Yi-Zhuang You.
\newblock Hamiltonian-driven shadow tomography of quantum states.
\newblock {\em Phys. Rev. Res.}, 4:013054, Jan 2022.

\bibitem{PRXQuantum.2.030348}
Senrui Chen, Wenjun Yu, Pei Zeng, and Steven~T. Flammia.
\newblock Robust shadow estimation.
\newblock {\em PRX Quantum}, 2:030348, Sep 2021.

\bibitem{2021arXiv210505992A}
Atithi {Acharya}, Siddhartha {Saha}, and Anirvan~M. {Sengupta}.
\newblock {Informationally complete POVM-based shadow tomography}.
\newblock {\em arXiv e-prints}, page arXiv:2105.05992, May 2021.

\bibitem{2021arXiv211002965L}
Ryan {Levy}, Di~{Luo}, and Bryan~K. {Clark}.
\newblock {Classical Shadows for Quantum Process Tomography on Near-term
  Quantum Computers}.
\newblock {\em arXiv e-prints}, page arXiv:2110.02965, October 2021.

\bibitem{PhysRevLett.127.110504}
Andrew Zhao, Nicholas~C. Rubin, and Akimasa Miyake.
\newblock Fermionic partial tomography via classical shadows.
\newblock {\em Phys. Rev. Lett.}, 127:110504, Sep 2021.

\bibitem{2022arXiv220713723W}
Kianna {Wan}, William~J. {Huggins}, Joonho {Lee}, and Ryan {Babbush}.
\newblock {Matchgate Shadows for Fermionic Quantum Simulation}.
\newblock {\em arXiv e-prints}, page arXiv:2207.13723, July 2022.

\bibitem{doi:10.1126/science.abk3333}
Hsin-Yuan Huang, Richard Kueng, Giacomo Torlai, Victor~V. Albert, and John
  Preskill.
\newblock Provably efficient machine learning for quantum many-body problems.
\newblock {\em Science}, 377(6613):eabk3333, 2022.

\bibitem{2022arXiv220203272B}
Kaifeng {Bu}, Dax {Enshan Koh}, Roy~J. {Garcia}, and Arthur {Jaffe}.
\newblock {Classical shadows with Pauli-invariant unitary ensembles}.
\newblock {\em arXiv e-prints}, page arXiv:2202.03272, February 2022.

\bibitem{2023PhRvA.107d2403K}
Jonathan {Kunjummen}, Minh~C. {Tran}, Daniel {Carney}, and Jacob~M. {Taylor}.
\newblock {Shadow process tomography of quantum channels}.
\newblock {\em \pra}, 107(4):042403, April 2023.

\bibitem{2023ScPP...14...94S}
Saumya {Shivam}, Curt~W. {von Keyserlingk}, and Shivaji~L. {Sondhi}.
\newblock {On classical and hybrid shadows of quantum states}.
\newblock {\em SciPost Physics}, 14(5):094, May 2023.

\bibitem{PhysRevLett.130.230403}
Matteo Ippoliti, Yaodong Li, Tibor Rakovszky, and Vedika Khemani.
\newblock Operator relaxation and the optimal depth of classical shadows.
\newblock {\em Phys. Rev. Lett.}, 130:230403, Jun 2023.

\bibitem{2022arXiv220912924B}
Christian {Bertoni}, Jonas {Haferkamp}, Marcel {Hinsche}, Marios {Ioannou},
  Jens {Eisert}, and Hakop {Pashayan}.
\newblock {Shallow shadows: Expectation estimation using low-depth random
  Clifford circuits}.
\newblock {\em arXiv e-prints}, page arXiv:2209.12924, September 2022.

\bibitem{2022arXiv221109835A}
Mirko {Arienzo}, Markus {Heinrich}, Ingo {Roth}, and Martin {Kliesch}.
\newblock {Closed-form analytic expressions for shadow estimation with
  brickwork circuits}.
\newblock {\em arXiv e-prints}, page arXiv:2211.09835, November 2022.

\bibitem{Akhtar:2022nme}
Ahmed~A. Akhtar, Hong-Ye Hu, and Yi-Zhuang You.
\newblock {Scalable and Flexible Classical Shadow Tomography with Tensor
  Networks}.
\newblock {\em Quantum}, 7:1026, 2023.

\bibitem{PhysRevResearch.5.023027}
Hong-Ye Hu, Soonwon Choi, and Yi-Zhuang You.
\newblock Classical shadow tomography with locally scrambled quantum dynamics.
\newblock {\em Phys. Rev. Res.}, 5:023027, Apr 2023.

\bibitem{koh2022classical}
Dax~Enshan Koh and Sabee Grewal.
\newblock Classical shadows with noise.
\newblock {\em Quantum}, 6:776, 2022.

\bibitem{Vovrosh:2021tc}
Joseph Vovrosh and Johannes Knolle.
\newblock Confinement and entanglement dynamics on a digital quantum computer.
\newblock {\em Scientific Reports}, 11(1):11577, 2021.

\bibitem{PhysRevResearch.3.043122}
Min Yu, Dongxiao Li, Jingcheng Wang, Yaoming Chu, Pengcheng Yang, Musang Gong,
  Nathan Goldman, and Jianming Cai.
\newblock Experimental estimation of the quantum fisher information from
  randomized measurements.
\newblock {\em Phys. Rev. Res.}, 3:043122, Nov 2021.

\bibitem{Noel:2022wr}
Crystal Noel, Pradeep Niroula, Daiwei Zhu, Andrew Risinger, Laird Egan,
  Debopriyo Biswas, Marko Cetina, Alexey~V. Gorshkov, Michael~J. Gullans,
  David~A. Huse, and Christopher Monroe.
\newblock Measurement-induced quantum phases realized in a trapped-ion quantum
  computer.
\newblock {\em Nature Physics}, 18(7):760--764, 2022.

\bibitem{koh2022experimental}
Jin~Ming Koh, Shi-Ning Sun, Mario Motta, and Austin~J. Minnich.
\newblock Experimental realization of a measurement-induced entanglement phase
  transition on a superconducting quantum processor, 2022.

\bibitem{doi:10.1126/science.aau4963}
Tiff Brydges, Andreas Elben, Petar Jurcevic, Beno{\^\i}t Vermersch, Christine
  Maier, Ben~P. Lanyon, Peter Zoller, Rainer Blatt, and Christian~F. Roos.
\newblock Probing renyi entanglement entropy via randomized measurements.
\newblock {\em Science}, 364(6437):260--263, 2019.

\bibitem{PRXQuantum.2.010307}
G.I. Struchalin, Ya.~A. Zagorovskii, E.V. Kovlakov, S.S. Straupe, and S.P.
  Kulik.
\newblock Experimental estimation of quantum state properties from classical
  shadows.
\newblock {\em PRX Quantum}, 2:010307, Jan 2021.

\bibitem{supp}
See Supplementary Material for: (i). Structures of deterministic layers; (ii).
  Gradient decent method of QNN's parameters. (iii). Training details of the
  observables learning task; (iv). Training details of the R\'enyi entropy
  measurement task; (v). Training details of the pattern recognition task;
  (vi). Comparison with fixed Computational Cost.

\bibitem{cerezo2021cost}
Marco Cerezo, Akira Sone, Tyler Volkoff, Lukasz Cincio, and Patrick~J Coles.
\newblock Cost function dependent barren plateaus in shallow parametrized
  quantum circuits.
\newblock {\em Nature communications}, 12(1):1791, 2021.

\bibitem{watrous_2018}
John Watrous.
\newblock {\em The Theory of Quantum Information}.
\newblock Cambridge University Press, 2018.

\bibitem{nielsen2002introduction}
Michael~A Nielsen.
\newblock An introduction to majorization and its applications to quantum
  mechanics.
\newblock {\em Lecture Notes, Department of Physics, University of Queensland,
  Australia}, 2002.

\bibitem{Uhlmann}
Peter Alberti and Armin Uhlmann.
\newblock Stochasticity and partial order. doubly stochastic maps and unitary
  mixing.
\newblock 01 1982.

\bibitem{Renyi3Poly}
The third order of single qubit reduced density matrix $tr[\hat{\rho}^3_a]$ is
  the second order of the elements of the density matrix.

\bibitem{netzer2011reading}
Yuval Netzer, Tao Wang, Adam Coates, Alessandro Bissacco, Bo~Wu, and Andrew~Y
  Ng.
\newblock Reading digits in natural images with unsupervised feature learning.
\newblock 2011.

\bibitem{Xiaoting}
Xiaokai Hou, Guanyu Zhou, Qingyu Li, Shan Jin, and Xiaoting Wang.
\newblock A duplication-free quantum neural network for universal
  approximation.
\newblock {\em Science China Physics, Mechanics \& Astronomy}, 66(7):270362,
  2023.

\bibitem{romero2017quantum}
Jonathan Romero, Jonathan~P Olson, and Alan Aspuru-Guzik.
\newblock Quantum autoencoders for efficient compression of quantum data.
\newblock {\em Quantum Science and Technology}, 2(4):045001, 2017.

\bibitem{bondarenko2020quantum}
Dmytro Bondarenko and Polina Feldmann.
\newblock Quantum autoencoders to denoise quantum data.
\newblock {\em Physical review letters}, 124(13):130502, 2020.

\bibitem{oh2020tutorial}
Seunghyeok Oh, Jaeho Choi, and Joongheon Kim.
\newblock A tutorial on quantum convolutional neural networks (qcnn).
\newblock In {\em 2020 International Conference on Information and
  Communication Technology Convergence (ICTC)}, pages 236--239. IEEE, 2020.

\bibitem{hur2022quantum}
Tak Hur, Leeseok Kim, and Daniel~K Park.
\newblock Quantum convolutional neural network for classical data
  classification.
\newblock {\em Quantum Machine Intelligence}, 4(1):3, 2022.

\bibitem{cong2019quantum}
Iris Cong, Soonwon Choi, and Mikhail~D Lukin.
\newblock Quantum convolutional neural networks.
\newblock {\em Nature Physics}, 15(12):1273--1278, 2019.

\bibitem{Caro:2022aa}
Matthias~C. Caro, Hsin-Yuan Huang, M.~Cerezo, Kunal Sharma, Andrew Sornborger,
  Lukasz Cincio, and Patrick~J. Coles.
\newblock Generalization in quantum machine learning from few training data.
\newblock {\em Nature Communications}, 13(1):4919, 2022.

\end{thebibliography}


\end{document}


\title{Supplementary Material for ''Randomness-enhanced Expressivity of Quantum Neural Networks''}

\author{Yadong Wu}
\affiliation{Department of Physics, Fudan University, Shanghai, 200438, China}
\affiliation{State Key Laboratory of Surface Physics, Key Laboratory of Micro and Nano Photonic Structures (MOE), Institute for Nanoelectronic Devices and Quantum Computing, Fudan University, Shanghai 200438, China}
\affiliation{Shanghai Qi Zhi Institute, AI Tower, Xuhui District, Shanghai 200232, China}

\author{Juan Yao}
\affiliation{Shenzhen Institute for Quantum Science and Engineering, Southern University of Science and Technology, Shenzhen 518055, Guangdong, China}
\affiliation{International Quantum Academy, Shenzhen 518048, Guangdong, China}
\affiliation{Guangdong Provincial Key Laboratory of Quantum Science and Engineering, Southern University of Science and Technology, Shenzhen 518055, Guangdong, China}

\author{Pengfei Zhang}
\email{pengfeizhang.physics@gmail.com}
\affiliation{Department of Physics, Fudan University, Shanghai, 200438, China}
\affiliation{Shanghai Qi Zhi Institute, AI Tower, Xuhui District, Shanghai 200232, China}

\author{Xiaopeng Li}
\email{xiaopeng$_$li@fudan.edu.cn}
\affiliation{Department of Physics, Fudan University, Shanghai, 200438, China}
\affiliation{State Key Laboratory of Surface Physics, Key Laboratory of Micro and Nano Photonic Structures (MOE), Institute for Nanoelectronic Devices and Quantum Computing, Fudan University, Shanghai 200438, China}
\affiliation{Shanghai Qi Zhi Institute, AI Tower, Xuhui District, Shanghai 200232, China}
\affiliation{ Shanghai Artificial Intelligence Laboratory, Shanghai 200232, China} 
\affiliation{Shanghai Research Center for Quantum Sciences, Shanghai 201315, China}

\maketitle

\section{Structures of deterministic layers}
\begin{figure}[h]
        \centering
        \includegraphics[width=0.3\linewidth]{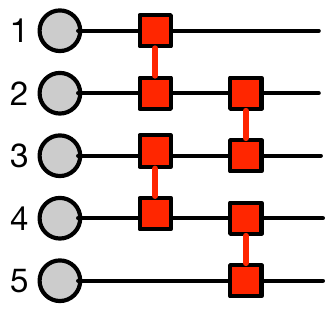}
        \caption{The brick wall structure of 5-qubit system. }
        \label{fig:BW}
    \end{figure}

In this section, we'll present the architecture of deterministic layers. The entire deterministic layer $\hat{U}$ is composed of a number of unites, i.e. $\hat{U}=\hat{V}^L\hat{V}^{l-1}\cdots\hat{V}^1$. Each unit $\hat{V}^a$ contains multiple two-qubit gates $\hat{v}_{ij}$ where $i,j$ represent the qubits indices. Each two-qubit gate is parameterized as $\hat{v}_{ij}=\exp(\sum_w \theta_{ij}^w\hat{g}_w)$ where $\{\hat{g}_k\}$ are the generators of the SU(4) group. Fig.[\ref{fig:BW}] illustrates the brick wall architecture of one unit. The explicit form of this single-unit neural network is as follow:
\begin{align}
\hat{V}=[\hat{\sigma}_0^1\otimes\hat{v}_{23}\otimes\hat{v}_{45}][\hat{v}_{12}\otimes\hat{v}_{34}\otimes\hat{\sigma}_0^5],
\end{align}
where $\hat{\sigma}_0$ is the identity of the single qubit. 
\vspace{5pt}

\section{Gradient decent method of QNN's parameters}
In this section, we present the calculation details of the gradients of parameters in QNNs with the random layer. The QNN architecture consists of three components: deterministic layers $\hat{U}_1$ and $\hat{U}_2$, a random layer $\mathcal{E}=\{w_i,\hat{U}_{r,i}\}$ situated between deterministic layers, and a classical function $f_{\bm{\beta}}$. The final prediction for the input state $|\psi_m\rangle$ is computed as follows:
\begin{align}
\label{FinalPre}
\mathcal{P}_m=\sum_{i=1}^{N_r} w_i \mathcal{P}_{i,m}=\sum_{i=1}^{N_r} w_i f_{\bm{\beta}}(\mathbf{p}_{i,m}).
\end{align}
where $\{w_i\}$ represents the probabilities of the random unitary operators in the ensemble $\mathcal{E}$, $\{\bm{\beta}\}$ are variational parameters in the classical function, and $\mathbf{p}_{i,m}$ denotes the measurements from the quantum circuit. In quantum machine learning tasks, the loss function $\mathcal{L}$ is a function of $\mathcal{P}_m$, and the gradient with respect to $\{w_i, \bm{\beta}\}$ can be easily obtained.
\begin{align}
\frac{\partial\mathcal{L}}{\partial w_i}&=\sum_m\frac{\partial\mathcal{L}}{\partial\mathcal{P}_m}\mathcal{P}_{i,m}\\
\frac{\partial\mathcal{L}}{\partial \bm\beta}&=\sum_m\frac{\partial\mathcal{L}}{\partial\mathcal{P}_m}\sum_i^{N_r}w_i\frac{\partial f_{\bm\beta}(\mathbf{p}_{i,m})}{\partial\bm\beta}
\end{align}

Each element of measurements is given by $p^s_{i,m}=\langle \psi_m| \hat{U}_1^\dagger\hat{U}_{r,i}^\dagger\hat{U}_2^\dagger\hat{M}_s\hat{U}_2\hat{U}_{r,i}\hat{U}_1|\psi_m\rangle$ where $\hat{M}_s$ is a general measurement operator. $\hat{U}_{r,i}$ is parametrized by $\{\bm \alpha_i^q\}$ that
\begin{align}
\hat{U}_{r,i}&=\hat{u}_1^i(\bm{\alpha}_{i}^{1})\otimes \hat{u}_2^i(\bm{\alpha}_{i}^{2})... \otimes \hat{u}_{N_{\text{sys}}}^i(\bm{\alpha}_{i}^{{N_{\text{sys}}}})\nonumber\\
\hat{u}_q^{i}&=\exp(\bm\alpha_i^q\cdot\bm\sigma),
\end{align}
where $\bm\alpha_i^q$ is a three-dimensional real vector and $\bm\sigma$ is a three-dimensional vector composed of the generators of the SU(2) group. The gradient of $\bm\alpha_i^q$ can be written as:
\begin{align}
\frac{\partial\mathcal{L}}{\partial \bm\alpha_i^q}&=\sum_m\frac{\partial\mathcal{L}}{\partial\mathcal{P}_m}\sum_i^{N_r}w_i\sum_s\frac{\partial f_{\bm\beta}(\mathbf{p}_{i,m})}{\partial p^s_{i,m}}\nonumber\\
\frac{\partial p_{i,m}^s}{\partial\bm\alpha_i^q}&=\langle \psi_m| \hat{U}_1^\dagger\hat{U}_{r,i}^\dagger\hat{U}_2^\dagger\hat{M}_s\hat{U}_2\frac{\partial\hat{U}_{r,i}}{\partial \bm\alpha_i^q}\hat{U}_1|\psi_m\rangle+h.c.\\
\frac{\partial\hat{U}_{r,i}}{\partial \bm\alpha_i^q}&=\hat{u}_1^i(\bm{\alpha}_{i}^{1})\otimes \hat{u}_2^i(\bm{\alpha}_{i}^{2})\cdots\otimes \frac{\partial\hat{u}_q^i}{\partial \bm\alpha_i^q} \otimes \cdots\hat{u}_{N_{\text{sys}}}^i(\bm{\alpha}_{i}^{{N_{\text{sys}}}})\nonumber
\end{align}
Then we can utilize the matrix exponential gradient \cite{PhysRevResearch.3.L032057} to calculate the gradient: $ \frac{\partial\hat{u}_q^i}{\partial \bm\alpha_i^q}$

Similarly, for the parameters $\{\bm\theta^l_{l_d}\}$ in the circuit of deterministic layers $\hat{U}_1,\hat{U}_2$, the gradient of parameters in this neural network can also be obtained from the matrix exponential gradient \cite{PhysRevResearch.3.L032057}.

\vspace{5pt}

\section{training details of observables learning task}

\subsection{Training Details}

In this section, we provide details about the observables learning task. While the quantum neural network contains $N_{\rm sys}$ qubits, the Hilbert space dimension is $\mathcal{D}=2^{N_{\rm sys}}$. Each input wave function $|\psi\rangle$ can be expanded as:
\begin{align}
\label{eq:wave}
|\psi\rangle=\frac{1}{\mathcal{N}}\sum_{s}^{\mathcal{D}}(a_s+ib_s)|s\rangle,
\end{align}
where $\{|s\rangle\}$ represents the bases of this Hilbert space, $\{a_s,b_s\}$ are randomly sampled from the uniform distribution $[-1,1]$, and $\mathcal{N}$ is the normalization factor. The target physical observables are diagonal in these bases with eigenvalues randomly sampled from the uniform distribution $[-2.5,2.5]$. To avoid overfitting, the size of the training set $N_{\mathcal{D}}$ must be larger than the degree of freedom of $\mathcal{D}$-dimensional hermition matrices, i.e. $N_{D}>\mathcal{D}^2=2^{2N_{\rm sys}}$. The output of the randomized neural network is $\mathcal{P}_m=\langle\psi_m|\sum_{i=1}^{N_r}p_i\hat{U}_{\textrm{tot},i}^\dagger(\beta_0\hat{I}+\beta_1\hat{\sigma}_z)\hat{U}_{\textrm{tot},i}|\psi_m\rangle$. We use Mean Square Error (MSE) as the loss function, denoted as $\mathcal{L}=\frac{1}{N_{\mathcal{D}}}\sum_m(\mathcal{P}_m-\mathcal{T}_m)^2$. To escape from the local minima, we use mini-batch Adam method to update parameters and set the learning rate $\eta=0.01$. Table [\ref{tab:training1}] shows the training dataset size for different quantum system size.

\begin{table}[h]
\renewcommand{\arraystretch}{2}
\caption{Hyper parameters of QNN}
\label{tab:training1}
\centering
\setlength{\tabcolsep}{5mm}{
\begin{tabular}{|c|c|c|c|c|c|}
\hline
$N_{\rm sys}$ & $\mathcal{D}$ &$L_2 $&$N_{D}$  &$N_{\rm batch}$&$\rm Epoch_{max}$\\
\hline
2&4&1&100&20&1000\\
\hline
3&8&4&200&40&1500\\
\hline
4&16&6&500&100&2000\\
\hline
5&32&8&1500&300&3000\\
\hline
\end{tabular}
}
\end{table}

\begin{figure}[h]
        \centering
        \includegraphics[width=0.5\linewidth]{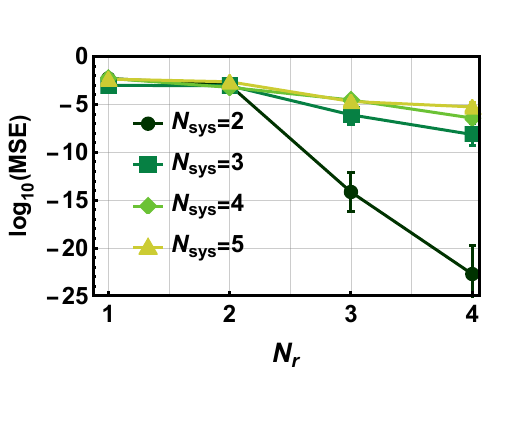}
        \caption{ Logarithmical mean square error of test dataset with 200 samples for $N_{\rm sys}=\{2,3,4,5\}$ and $N_r=\{1,2,3,4\}$. Markers are the averaged loss of 10 different training processes and error bars are the standard deviation. }
        \label{fig:obstest}
    \end{figure}

We also apply the optimized neural network to the testing dataset containing $N_{\rm test}=200$ samples. Fig.[\ref{fig:obstest}] shows the logarithmical mean square error of test dataset for different system size.
\vspace{5pt}

\subsection{Opening the black box}
Based on the training results presented in the main text, it was found that for the two-qubit system, by setting $N_r=3$, the loss can be significantly reduced as low as $10^{-15}$.  We further investigate the effectiveness of the quantum neural network in learning observable expectations. As shown in Fig.[2] in the main text, when $N_{\rm sys}=2$, $N_r=3$, we can predict the observable's expectation with extremely high accuracy. After training, we find that the probabilities of each random unitary are almost same $w_i=1/N_r$. Consequently, the final prediction of the neural network is $\mathcal{P}=\frac{1}{3}\sum_{i=1}^{N_r=3}\mathcal{P}_i$ where $\mathcal{P}_i=\langle\psi|\hat{U}_{\textrm{tot},i}^\dagger(\beta_0\hat{\sigma}_0+\beta_1\hat{\sigma}_z)\hat{U}_{\textrm{tot},i}|\psi\rangle$. This results in the predicted operator reading as:  
\begin{align}
\hat{O}_{\rm pre}=\frac{1}{3}\sum_{i=1}^{N_r=3}\hat{o}_i,~~~~~
\end{align}
with $\hat{o}_i=\hat{U}^\dagger_{\textrm{tot},i}(\beta_0\hat{\sigma}_0+\beta_1\hat{\sigma}_z)\hat{U}_{\textrm{tot},i}$. Noticing that the target observable is diagonal, it can be found that, after training, each element of $\hat{O}_{\rm pre}$ is very close to that of $\hat{O}$. 

We expand each predicted operator $\hat{o}_i$ in Pauli string operator bases $\hat{o}_{i}=\sum_{ab}C^i_{ab}\hat{\sigma}_a\otimes\hat{\sigma}_b$ where $a,b=x,y,z,0$. $C_{00}$ represents the trace of $\hat{O}$ and $C_{zz},C_{z0},C_{0z},C_{00}$ can reconstruct the diagonal elements of $\hat{O}_{\rm pre}$. The random unitaries are tensor products of unitaries applied to each single-qubit, i.e. $\hat{U}_{r,i}=\hat{u}_1^i\otimes\hat{u}_2^i$. For the sake of simplicity, $\hat{U}_{r,1}$ can be absorbed into $\hat{U}_2$. We then found that $\hat{U}_{r,2}$ and $\hat{U}_{r,3}$ commute with the Pauli operator $\hat{\sigma}_z\otimes\hat{\sigma}_0$ and $\hat{\sigma}_0\otimes\hat{\sigma}_z$. This implies that $\hat{U}_{r,2}$ and $\hat{U}_{r,3}$ rotate each qubit in its own Bloch sphere along $z$ axis. So from the coefficients obtained through expansion, we can obtain the angles projected in the $x-y$ plane:  
\begin{align}
\xi_{1,z}&=\arctan(C_{zy}/C_{zx}), ~~~\xi_{1,0}=\arctan(C_{0y}/C_{0x}),\\
\xi_{2,z}&=\arctan(C_{yz}/C_{xz}), ~~~\xi_{2,0}=\arctan(C_{y0}/C_{x0}).
\end{align}
For $N_r=3$, the angle difference of each qubit between different $\hat{o}_i$ is $2\pi/3$. This implies that $\xi_{l,a}^1-\xi_{l,a}^2=\xi_{l,a}^2-\xi_{l,a}^3= \xi_{l,a}^3-\xi_{l,a}^1=2\pi/3$ where $l=\{1,2\},~a=\{z,0\}$. Furthermore, table[\ref{tab:coeO}] illustrates certain relations between coefficients, ensuring the summation of off-diagonal elements of $\hat{o}_i$ vanishes.

\begin{table}[h]
\renewcommand{\arraystretch}{2}
\caption{Relations between coefficients of predicted operators}
\label{tab:coeO}
\centering
\setlength{\tabcolsep}{4mm}{
\begin{tabular}{|c|c|c|c|c|}
\hline
\diagbox{a}{b}& x & y & z &0\\
\hline
x&$C_{xx}$&$C_{xy}$&$C_{xz}$&$C_{xz}$\\
\hline
y&$-C_{xy}$&$C_{xx}$&$C_{yz}$&$C_{yz}$\\
\hline
z&$C_{zx}$&$C_{zy}$&$C_{zz}$&$C_{z0}$\\
\hline
0&$C_{zx}$&$C_{zy}$&$C_{0z}$&$C_{00}$\\
\hline
\end{tabular}
}
\end{table}

\section{Training details of R\'enyi entropies learning task}

\subsection{Training Details }
In this section, we'll describe the data generation process for the R\'enyi entropies learning task and the optimization method for parameters. Similar to the observable learning task, we consider a quantum system containing $N_{\rm sys}=5$ qubits. The Hilbert space dimension is $\mathcal{D}=2^5$ and each input wave function is generated as given by  eq. (\ref{eq:wave}). After the evolution, we measure the subsystem in the bases $x_s=|\langle s|\hat{U}_{r,i}|\psi\rangle|^2,~ s=1,2,\cdots,2^{N_{\rm sub}}$. When learning the purity, $\mathcal{T}={\rm Tr}[\hat{\rho}_{\rm sub}^2]$ is a quadratic function of reduced density matrices. Therefor, in this case we set the classical non-linear function to also be a quadratic function of measurement results $\mathbf{x}$.
\begin{align}
f_{\boldsymbol{\beta}}(\vec{x})=\beta_0+\vec{\beta}_1^T\vec{x}+\vec{x}^T\boldsymbol{\beta}_2\vec{x}
\end{align}

We also use mini-batch Adams to optimize parameters in these randomized quantum neural networks. Specifically, we set $N_{D}=100,~N_{\rm batch}=20$ for $N_{\rm sub}=1$ and $N_{D}=200,~N_{\rm batch}=40$ for $N_{\rm sub}=2$. The learning rate is set to $\eta=0.01$. After training, we apply this neural network to a test dataset consisting of $N_{\rm test}=200$ samples to ensure that the neural network has learnt the purities.

\subsection{Convergence of Randomness for n=2 R\'enyi Entropy}
\begin{figure}
       \centering
        \includegraphics[width=0.95\linewidth]{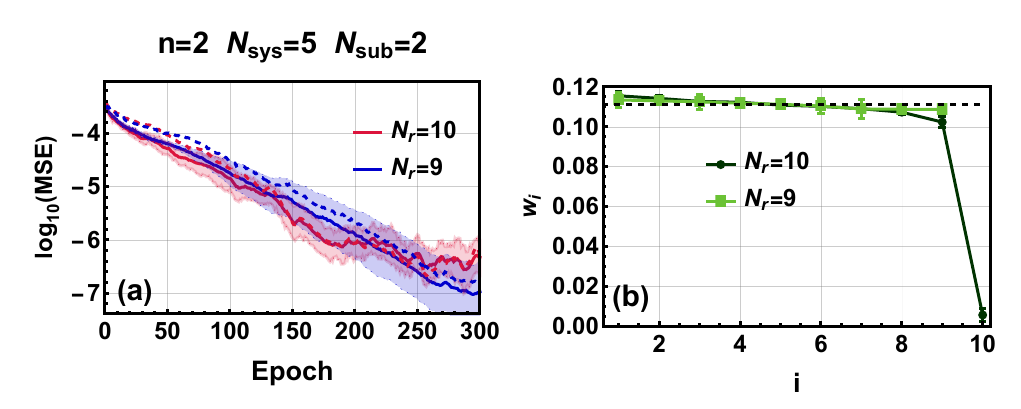}
        \caption{  The results for learning the second order Tr$[\hat{\rho}_{\rm sub}^2]$ for $N_{\rm sub}=2$. (a) The logarithmical mean square error for $N_r=9$ and $N_r=10$ is averaged over the training process for 10 different random initializations, with the shaded region representing the standard deviation. The dashed lines represent the validation loss using a dataset containing 200 samples. (b) Probabilities of random unitary operators for $N_r=9$ and $N_r=10$ in decreasing order. These optimal parameters $\{w_i\}$ are averaged over the training process for 10 different random initializations with error bars included.}
        \label{fig:difNr}
    \end{figure}

In the main text, we demonstrated that for $N_{\rm sub}=1$, we need $N_r=3$ random unitary operators, and for $N_{\rm sub}=2$, we require $N_r=9$ random unitary operators to predict $n=2$ R\'enyi entropy with high accuracy. Here, for $N_{\rm sub}=2$, we further support our conclusion by comparing the learning performance of $N_r=9$ and $N_r=10$. Both loss functions decay rapidly with exponential behavior and converged to nearly the same value, as shown in Fig. \ref{fig:difNr}(a). After training, we extracted the probabilities $\{w_i\}$ of each random unitary operator. They are all approximately equal to $1/9$ for $N_r=9$, and one probability would vanish for $N_r=10$. This implies that for $N_{\rm sub}=2$, $N_r=3^2$ is enough for achieving good performance.

\subsection{Third Order Entropy Learning}

We also applied the randomized quantum neural networks to learn higher-order R\'enyi entropies, specifically, $\mathcal{T}=\textrm {Tr}[\hat{\rho}_{\rm sub}^3]$. In this case, the label is a third order polynomial function of the density matrix. As a result, we set the classical non-linear function up to be a third-order polynomial of measurement results:
\begin{align}
f_{\boldsymbol{\beta}}(\vec{x})=\beta_0+\vec{\beta}_1^T\vec{x}+\vec{\beta}_2^T\vec{x}^{\otimes2}+\vec{\beta}^T_3\vec{x}^{\otimes3}
\end{align}
However, after training, we observed that the third-order coefficients $\vec{\beta}_3$ vanish for the subsystem with one qubit. The loss for $N_r=3$ also decreases exponentially which is consistent with the analytical calculation for a general single-qubit density matrix:
\begin{align}
\hat{\rho}_{\rm sub}&=\left(
\begin{matrix}
\rho_{11}&\rho_{12}^r+i\rho_{12}^i\nonumber\\
\rho_{12}^r-i\rho_{12}^i&1-\rho_{11}\label{eq:tr3}
\end{matrix}
\right)\\
\textrm{Tr}[\hat{\rho}^3_{\rm sub}]&=1-3\rho_{11}+3\rho_{11}^2+3| \rho_{12}|^2
\end{align}
where $\rho_{12}=\rho_{12}^r+i\rho_{12}^i$, and $\hat{\rho}_{\rm sub}$ is semi-positive with unit trace. The single-qubit R\'enyi entropies with $n=2,3$ have a similar form. $\textrm{Tr}[\hat{\rho}^2_{\rm sub}]=1-2\rho_{11}+2\rho_{11}^2+2| \rho_{12}|^2$. Therefore, $N_r=3$ is also suitable for $n=3$ R\'enyi entropy. However, for $N_{\rm sub}=2$, $\vec{\beta}_3$ makes a great contribution.As shown in Fig. 3(d) in the main text, increasing $N_r$ leads to the loss converging to a lower value.

\section{Training Details of Pattern Recognition Task}

\subsection{Training Details}

In this section, we'll provide details on how we encode the "Street View of House Number (SVHN)" data into the 5-qubit quantum system, explain the explicit form of the nonlinear function, and describe the optimization methods used.

Each pattern in the original SVHN dataset contains a grid of $32\times32$ pixels, with each pixel having three RGB channels. Initially, we convert these RGB images to grayscale using a weighted combination: $\rm 0.2989 R+0.5870G+0.1140B$. Next,  we resize the $32\times32$ pixel images to $8\times8$ pixels by averaging the nearest 4-by-4 neighborhood. Then, we reshape this resulting $8\times8$ matrix into a $32\times2$ matrix, with the first column representing the real part of the coefficients of the basis and the second column representing the imaginary part of the coefficients in eq.(\ref{eq:wave}). For demonstration, we selected two patterns from all the images.  We labeled one pattern as '1' and the other as '0'. After measuring the single qubit, we set the non-linear function as a polynomial of the measurement output $x$:
\begin{align}
f_{\boldsymbol{\beta}}(x)=\sum_{k=0}^5\beta_kx^k
\end{align}
And for this classification task, we take cross-entropy as the loss function:
\begin{align}
\mathcal{L}=\frac{1}{N_{D}}\sum_m^{N_{D}}-\mathcal{T}_m\log(\mathcal{G}_m)-(1-\mathcal{T}_m)\log(1-\mathcal{G}_m)
\end{align}
where $\mathcal{G}_m=1/1+\exp(-\mathcal{P}_m)$ in the region (0,1). We also use mini-batch Adams to optimize parameters in this randomized quantum neural network with $N_{D}=1200, L_1=L_2=4,N_{\rm batch}=300,\eta=0.01$.

\section{ Comparison with fixed Computational Cost}
In this section, we discuss the computational cost of the QNN with a random layer. In our approach, we introduce one random layer into the QNN, and the final prediction is based on the ensemble average of this random layer. The computational cost of the QNN scales linearly with both $N_r$ and the depth of the deterministic layers $L_{1}, L_2$. Therefore, to make a fair comparison, we consider the results of different $N_r$ with the same $N_r(L_1+L_2)$.

\begin{figure}
       \centering
        \includegraphics[width=0.95\linewidth]{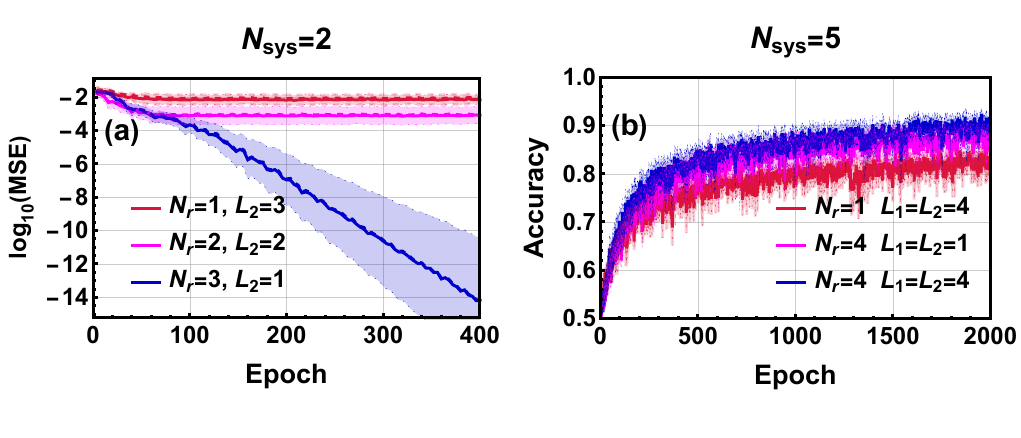}
        \caption{  Training results comparison with the same computational cost. (a) The observable learning result for $N_{\rm sys}=2$ with save $N_rL_2$. (b) The patter recognition result for $N_{\rm sys}=5$. The red line and the pink line contain same computational cost $N_r(L_1+L_2)=8$. The training results are averaged over the training process for 10 different random initializations, and the shaded region represents the standard deviation. The dashed lines are the validation loss with the dataset containing 200 samples.}
        \label{fig:fixC}
    \end{figure}

 For the observable learning task we set the deterministic layer $\hat{U}_1$ as the identity, i.e. In this context, using $L_1=0$. $N_r=3, L_2=1$ is sufficient to predict the observables' expectations with high accuracy for a quantum system with $N_{\rm sys}=2$. In our investigation, we set $L_2=3,2$ for $N_r=1,2$ while keeping $N_rL_2$ approximately same. FIG. \ref{fig:fixC}(a) illustrates that, for the same computational cost, introducing appropriate randomness can enhance the expressivity of the QNN. This enhancement occurs without the necessity of increasing the depth of the QNN. Even when considering $N_r=2, L_2=2$, the computational cost is higher than when using $N_r=3, L_2=1$.
 
Similar results are observed for the pattern recognition task. We set the deterministic layers $\hat{U}_1, \hat{U}_2$ as variational with the same circuit depth $L_1=L_2$.  Pattern recognition is a typical machine learning task, where the both data and the function are classical. Thus FIG.\ref{fig:fixC}(b) shows that, even with just one brick wall unit in the deterministic, the configuration with $N_r=4$ also has better learning performance than that of $N_r=1,L_1=L_2=4$.

\bibliographystyle{unsrt}  
